\documentclass[twocolumn,showpacs,preprintnumbers,amsmath,amssymb]{revtex4}

\usepackage{graphicx}
\usepackage{dcolumn}
\usepackage{bm}

\begin{document}

\title{Direction Detector on an Excitable Field: Field Computation with Coincidence Detection}


\author{Hiroki Nagahara}
\affiliation{Department of Physics, Graduate School of Science, Kyoto University \& CREST, Kyoto 606-8502, Japan}

\author{Takatoshi Ichino}
\altaffiliation{Present Address: Venture Business Laboratory, Kyoto University, Kyoto 606-8501, Japan.}
\affiliation{Department of Physics, Graduate School of Science, Kyoto University \& CREST, Kyoto 606-8502, Japan}

\author{Kenichi Yoshikawa}
\email[Author to whom correspondence should be addressed. Electronic address: ]{yoshikaw@scphys.kyoto-u.ac.jp}
\affiliation{Department of Physics, Graduate School of Science, Kyoto University \& CREST, Kyoto 606-8502, Japan}


\date{\today}

\begin{abstract}

Living organisms process information without any central control unit and without any ruling clock.
We have been studying a novel computational strategy that uses a geometrically arranged excitable field, i.e., ``field computation."
As an extension of this research, in the present article we report the construction of a ``direction detector" on an excitable field.
Using a numerical simulation, we show that the direction of a input source signal can be detected by applying the characteristic as a ``coincidence detector" embedded on an excitable field.
In addition, we show that this direction detection actually works in an experiment using an excitable chemical system.
These results are discussed in relation to the future development of ``field computation."

\end{abstract}


\maketitle


\section{Introduction}
Modern digital computers process information in a step-wise manner using a CPU equipped with a ruling clock.
Improvements in the step-wise operation, with regard to speed and reliability, have made digital computers indispensable in modern society.
However, step-wise operation under the control of a CPU inevitably exhibits a serious bottleneck when performing ``real-time operations" on large sets of time-dependent input data.
To break through such a bottleneck, many approaches to ``parallel computation" have been proposed.
Most previous studies on ``parallel computation" seem to have focused on the problem of how to effectively connect multiple unit computers (see, e.g., \cite{NStndCmptr}).
On the other hand, living organisms routinely perform ``parallel computation" in the absence of any ruling clock.
Higher order animals obtain the information from the external world through sensory systems, make decisions, and act optimally, in ``real time" and in the proper order.
Thus, an investigation of ``parallel computation" that mimics the essentials of computation in living organisms may be of scientific value.

It is well known that characteristics of excitability can be found in a wide variety of biological systems, including nerve membranes, and that living organisms use excitability in cells or tissue for information processing \cite{GmtryBTm, MthBio, MthPhys}. 
Studies on information processing in excitable media may lead to a fundamental understanding of information processing in biological systems. 
Several studies have provided interesting examples of information processing with an excitable field, such as in image processing \cite{Kuhnert1986, KAK,Rambidi2002, NIM} and logic operations \cite{TSh, SKSh, Sielewiesiuk2001}.
Recently, it has also been reported that, in both a simulation \cite{KYAAOHY,ADK,Mtk,MIYN2001,SCSG2002} and an actual experiment on an excitable system \cite{AAYY1996, Icn}, ``diode" can be made to appear between opposing excitable fields by choosing the spatial arrangement of excitable fields in an asymmetric manner.

By applying the unique characteristics of such diodes on an excitable field, logic gates with the operations AND, OR, and NOT can be created \cite{Mtk}, where the effect of the coincidence of a pair of signals with a time difference plays an essential role.
Such logic gates naturally lead to ``time-dependent" signals in the absence of any timing apparatus, i.e., without a ``clock."
As an extension of these logic gates, interesting examples of information processing have been reported such as a time-interval detector, a comparison detector, and a subtraction detector\cite{MY}.
It has also been reported that, with excitable fields in a comb-like arrangement, the coincidence or time-difference detector of a couple of input pulses can also be constructed in a real experiment using an excitable chemical system  \cite{Icn}, the Belousov-Zhabotinsky (BZ) reaction \cite{Zakin1970,Winfree1972}, and in computer simulations \cite{Mtk}.
The detection of a time-difference between a pair of input pulses is one of the basic functions of information processing in nervous systems, such as in sound localization in auditory systems \cite{CM,GRT}.

In this study, by extending the idea of a ``coincidence detector," we report a simulation, and a corresponding actual experiment, of a ``direction detector" that can detect the direction from which a chemical wave propagates, using an excitable field with a suitable geometric arrangement.

\section{Numerical Methods}
As an actual experimental system, in the present study we adopted the photosensitive BZ reaction.
In this chemical system, illumination with light results in the production of bromide, which inhibits the oscillatory reaction.
Thus, the excitability can be adjusted by changing the intensity of illumination \cite{Kuhnert1986,ADK,KATK}.
The essential features of the photosensitive BZ system can be described by a two-variable Oregonator model \cite{Field1973, Tyson1980} that has been modified to include the photochemical pathway \cite{HPL, TSPK}, as in Eq. \ref{Eq:MO2}:
		\begin{align}
			\frac{\partial u}{\partial t}&=D_u \nabla^2u + \frac{1}{\epsilon}\left\{ u(1-u) -(fv+\phi)\frac{u-q}{u+q} \right\}, \notag\\
			\frac{\partial v}{\partial t}&=u-v,\label{Eq:MO2}
		\end{align}
where the variables $u$ and $v$ correspond to the dimensionless concentrations of activator and inhibitor, respectively.
$D_u$ is the diffusion coefficient of the activator.
The parameter $\phi$ corresponds to the light intensity, and the local excitability of the medium is inhibited when $\phi$ is high \cite{MSCS}.
The kinetic parameters $\epsilon=0.05$, $q=0.00015$, $f=1.0$ and $D_u=1.0$ are taken to be constant throughout the simulation. 
The medium is excitable at $\phi =0.007$ and becomes nonexcitable as $\phi$ increases.
Numerical simulations were carried out with Eq. \ref{Eq:MO2} using the ADI method (alternating direction implicit method), with differential calculus and the Runge-Kutta method\cite{NumRcp}.
The boundary condition at the edge of the frame is taken to be no flux (Neumann condition), while the flux between the excitable and inhibitory fields is taken to be free.
The grid size is $300\times300$ points in a square lattice, and the time interval is taken to be $\Delta t=0.0001$.
The mesh size is taken to be $0.2$.

\section{Numerical Results}
\begin{figure}[h]
\includegraphics[width=90mm]{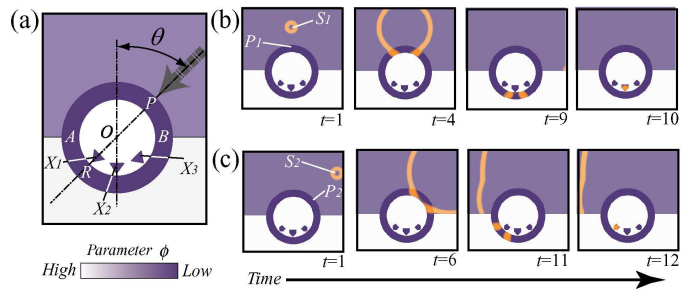} 
\caption[Numerical result]{\label{fig:NumRslt1}(a)Schematic representation of a ``direction detector" on an excitable field.
The shallow and deep purple regions correspond to excitable fields, while the white region corresponds to an inhibitory field.
The field geometry is arranged so that a signal wave is transmitted into one of the output channels $X_1-X_3$ depending on the direction that a wave triggered in the shallow purple area propagates.
(b)An example of a propagating wave.
A wave triggered at point $S_1$ causes a signal to be transmitted to output channel $X_2$.
(c)Another example of a propagating wave.
A wave triggered at the point $S_2$ causes a signal to be transmitted to output channel $X_1$.}
\end{figure}

Figure \ref{fig:NumRslt1} depicts a ``direction detector" constructed of excitable and inhibitory fields.
In Fig. \ref{fig:NumRslt1}(a), both the deep and shallow purple regions correspond to excitable fields ($\phi = 0.007$ and $\phi=0.033$ in Eq. \ref{Eq:MO2}), while the white region corresponds to an inhibitory field ($\phi =0.209$).

Figures \ref{fig:NumRslt1}(b) and \ref{fig:NumRslt1}(c) show time series of the propagation of a signal wave.
In Fig. \ref{fig:NumRslt1}(b), a wave triggered at point $S_1$ (Fig. \ref{fig:NumRslt1}(b) $t=$1) enters the ring-shaped excitable field through the point $P_1$ and splits into a pair of rotating pulses (Fig. \ref{fig:NumRslt1}(b) $t=$4), which causes the signal to be transmitted to output channel $X_2$ (Fig. \ref{fig:NumRslt1}(b) $t=$10).
On the other hand, in Fig. \ref{fig:NumRslt1}(c), a wave triggered at point $S_2$ (Fig. \ref{fig:NumRslt1}(c) $t=$1) enters the excitable ring through point $P_2$ and splits into a pair of rotating pulses (Fig. \ref{fig:NumRslt1}(c) $t=$6), which causes the signal to be transmitted to output channel $X_1$.
Thus, the transmission of the signal onto an output channel is generated only if the pair of pulses collide with each other near the gate to the output channel.
It is clear that the output channels $X_1$-$X_3$ provide the information on the original direction of the input signal.

\section{Experimental Methods and Results}
\begin{figure}[h]
\includegraphics[width=90mm]{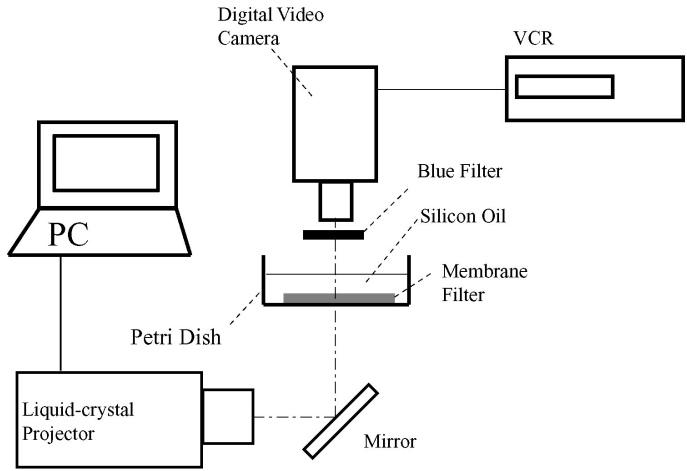}
\caption[Setup]{\label{fig:ExpSetup}Scheme of the experimental setup.}

\end{figure}

We examined the above idea in a real experiment.
As an experimental model, we adopted a photosensitive version of the BZ reaction under conditions similar to those in previous studies \cite{Icn, KATK}.
The water was purified with a Millipore-Q system.
The BZ solution we used in this experiment consisted of $\rm{6.8 \times 10^{-1}M}$ sodium bromate ($\rm{NaBrO_3}$), $\rm{3.0\times 10^{-1}M}$ sulfuric acid ($\rm{H_2SO_4}$), $\rm{2.0\times 10^{-1}M}$ malonic acid ($\rm{CH_2(COOH)_2}$), $\rm{5.0\times 10^{-2}M}$ sodium bromide ($\rm{NaBr}$), and $\rm{1.7\times 10^{-3}M}$ ruthenium(II)-bipyridyl ($\rm{Ru(bpy)_3^{2+}}$).
Cellulose-nitrate membrane filters (Advantec, A100A025A) with a pore size of 1$\mu m$ were soaked in BZ solution for about 1 minute.
The membrane was gently wiped with filter paper to remove excess water and placed in a petri dish.
The surface of the membrane filter was immediately covered with silicon oil (Shin-Etsu Chemical Co.) to prevent it from drying and to protect it from the influence of oxygen.
The experiments were carried out in an air-conditioned room at $21\pm1^{\circ}C$, at which the reaction medium showed no spontaneous excitation.
We created a geometry composed of white, gray, and black regions, similar to the circuit in Fig. \ref{fig:NumRslt1}(a), with a personal computer.
Such image was illuminated on the medium from the below through a liquid-crystal projector (Epson, ELP-810) as shown schematically in Fig. \ref{fig:ExpSetup}.
The liquid-crystal projector was used as the light source, and the spatial intensity distribution was controlled by the computer.
The experiments were monitored from above with a digital video camera (Panasonic, NV-DJ100).
For image enhancement, a blue optical filter (AsahiTechnoGlass, V-42) with a maximum transparency at 410 nm was used.
The light intensity at the illuminated area was determined by a light intensity meter (ASONE, LM-332).

\begin{figure}
\includegraphics[width=90mm]{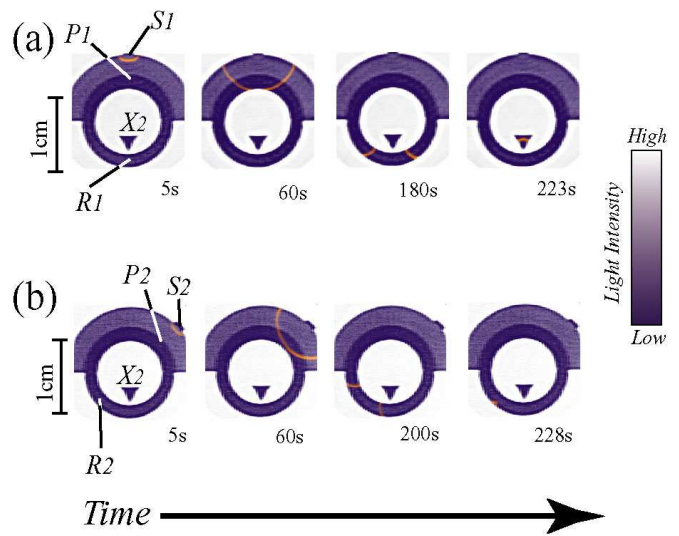}
\caption[Result of Ex A]{\label{fig:ExpResult}Experimental results with a ``direction detector" using the photosensitive BZ reaction. This detector is the simplest version with only one output channel ($X_2$). In order to show the result as an illustration manner, quasi-color is used in the figure. (a)A signal wave is transmitted to the output channel $X_2$ when a wave propagates from above. (b)No signal wave is transmitted to output channel $X_2$ when a wave does not propagate from above.
 
}
\end{figure}

Figure \ref{fig:ExpResult} shows the experimental results in the photosensitive BZ reaction, where the white and purple regions correspond to inhibitory and exciteble fields, respectively.
In Fig. \ref{fig:ExpResult} the white, shallow purple, and deep purple regions are illuminated at the intensity of $6.4\times 10^4$lx, $6.6\times 10^3$lx, and $2.0\times 10^3$lx, respectively.
Chemical waves are colored orange.
This detector is made as simple as possible, so it has only one output channel ($X_2$).
In Fig. \ref{fig:ExpResult}(a), a chemical wave triggered at the point $S_1$ propagates into the ring-shaped excitable field through point $P_1$, which is the nearest to $S_1$ on the excitable ring.
The wave then splits into two pulses that undergo clockwise and counterclockwise rotation, respectively.
The two pulses collide with each other and disappear at point $R_1$, which is opposite $P_1$, and a signal is transmitted to output channel $X_2$ (Fig. \ref{fig:ExpResult}(a) 223s).
In a similar way, in Fig. \ref{fig:ExpResult}(b), a chemical wave triggered at $S_2$ enters the ring through $P_2$.
The wave then splits into two pulses that undergo clockwise and counterclockwise rotation and collide with each other at point $R_2$, which is opposite $P_2$ (Fig. \ref{fig:ExpResult}(b) 228s).
No signal is transmitted to output channel $X_2$, even though the clockwise pulse passes near the gap between $X_2$ and the excitable ring.
Thus, the essential characteristics of the ``direction detector" in the numerical simulation were realized in the experiment.

\section{Discussion and Conclusion}
We have shown that, in both a simulation and an actual experiment on an excitable system, a ``direction detector" can be created by suitbly arranging the spatial geometry of excitable fields.
Although we adopted the Oregonator (Eq. \ref{Eq:MO2}) for an excitable field in the numerical simulation in our present study, the essential feature of wave propagation on an excitable media can also be described by the FitzHugh-Nagumo equation \cite{NGM62,FH61}, since both are reaction-diffusion equations.
In our previous studies, we have shown various kinds of ``field computations" in numerical simulations using FitzHugh-Nagumo-type equation \cite{Mtk,MY}.

In the ``direction detector" as Fig. \ref{fig:NumRslt1}(a), the location of the azimuth is converted into a timing difference between gates $A$ and $B$.
Next, through the AND gates (logic product) at the output $X_k$ ($k=1, 2, 3$), this timing difference is detected as the position of the collision.
Thus, part of the mathematical structure of the ``direction detector" can be easily described by using a function as in Eq. \ref{Eq:hevi}.
\begin{align}
\chi_{[0,\alpha]}(x)=\left\{
	\begin{array}{lcr}
	0&  (x\notin [0,\alpha]),\\
	1&	(x\in [0,\alpha]).\label{Eq:hevi}
	\end{array}
	\right.
\end{align}

Let us suppose in Fig. \ref{fig:NumRslt1}(a) 
 , after a wave is triggered at a certain time and in a certain direction $\theta$, two pulses are triggered at the points $A$ and $B$ at time $t_1$ and $t_2$, respectively.
That is, the location of the azimuth $\theta$ is converted into information in the two pulses which travel along the ring-shaped excitable field with a time difference.
If we associate the arrival of one pulse with an input of logical true, then the input to gate $X_k$ $(k=1,2,3)$ by the counterclockwise pulse is described by $\chi_{[0,\alpha]} \left(c(t-t_1)-\pi d k/n \right)$, where $d=AO=BO$, $c$ is the propagation speed of the pulses on the excitable ring, and $n$ is the number of the segments divided by the output channels $X_k$, i.e., the number of output channels $+1$ ($n=4$ in Fig. \ref{fig:NumRslt1} (a)).
$\alpha$ is a positive constant and the interval $[0, \alpha]$ indicates the permissible range of the time difference: the excitability of the field and the shape of output gates $X_k$ are reflected in this parameter.
In the same way, the input to gate $X_k$ by the clockwise pulse is written as $\chi_{[0,\alpha]} \left(c(t-t_2)+{\pi d}(k-n)/n \right)$.
The combination of these two representations with a logical product (AND) yields the following representation:
\begin{align}
f_k(t;t_1,t_2)=&\chi_{[0,\alpha]} \left(c(t-t_1)-\frac{\pi d}{n}k \right)\notag\\
				&\cdot\chi_{[0,\alpha]} \left(c(t-t_2)+\frac{\pi d}{n}(k-n) \right).\label{Eq:thethe}
\end{align}
The function $f_k(t; t_1, t_2)$ corresponds to the logical value of the output $X_k$.
$f_k(t; t_1, t_2)=1$ means that the pulses collide with each other near gate $X_k$, and $f_k(t;t_1, t_2)=0$ means that the collision does not happen near gate $X_k$.
In the representation in Eq. \ref{Eq:thethe}, we have done a reduction and neglected the pair annihilation of the pulses.
Note that Eq. \ref{Eq:thethe} is a logical expression that explicitly includes the continuous variable of time $t$.

As we stated above, the ``direction detector" in Fig. \ref{fig:NumRslt1} uses a time difference between two signals with no discretization of time by a ruling clock.
With the use of such a ``time-logic operation," it should be possible to construct a novel type of parallel computing machine in the absence of a ruling clock.
For a Neumann-type computer equipped with a CPU to deal with such ``time-dependent" signals, time information has to be discretized, recorded on a one-dimensional matrix and processed in a step-wise manner.
On the other hand, living organisms continuously process many signals without any governing clock, and signals with time-sequential information are encoded in wave packets or successive pulses.
It was reported that a type of owl uses an interaural time difference (ITD) to detect the direction of the sound made by its prey \cite{CM, GRT}.
Therefore, the time operation on an excitable field may more closely resemble the information processing in biological systems than Neumann-type computation.

As we have described above, the promising properties of ``field computation" can be realized in a chemical reaction system, i.e., the BZ reaction.
However, the problem with the BZ reaction is its fragility; the liquid reaction medium is not suitable for use in a practical computing machine. 
Thus, ``field computation" that takes advantage of time-sequential information could be implemented in a solid circuit by using modern electronic technology, as in semiconductor production \cite{Amemiya2002} or in a solid surface \cite{Grossen1995}.
To overcome this fragility, it may be necessary to expand the essential idea of ``field computation."
Recent advances in computer development have involved the fabrication of networks, whose characteristics can be described by coupled differential equations or finite-difference equations without an explicit inclusion of real space.
However, due to the many effects of miniaturization, it may be necessary to explicitly consider spatial extension at the individual nodes in a network.
In other words, we have to describe the electronic network with partial differential equations instead of coupled ordinary differential equations.
Thus, the novel idea of ``field computation" may offer a path for the future development of ``computers," by avoiding a serious bottleneck in current Neumann-type computation.

\begin{acknowledgments}
We would like to thank Dr. Ikuko N. Motoike (Future University-Hakodate, Japan)  for the precious comments and discussions.
\end{acknowledgments}


\newpage

\end{document}